\documentclass[twocolumn]{aastex61}
\title{1I/2017 U1 Rotation Paper}

\newcommand\aastex{AAS\TeX}

\newcommand{\Ou}{`Oumuamua}
\usepackage{xcolor}
\usepackage{url}

\accepted{ApJ Letters 865 L21 (2018)}

\shorttitle{\aastex\ 1I/2017 U1 Rotation}
\shortauthors{Belton et al.}

\begin{document}

\title{The Excited Spin State of 1I/2017 U1 `Oumuamua}

\correspondingauthor{Michael J. S. Belton}
\email{mbelton@dakotacom.net}

\author[0000-0002-1542-1903]{Michael J. S. Belton}
\affiliation{Belton Space Exploration Initiatives, LLC, 430 Randolph Way, Tucson AZ 85716 USA}
\affiliation{Kitt Peak National Observatory, Tucson, AZ 85719, USA}

\author[0000-0001-6952-9349]{Olivier R. Hainaut}
\affiliation{European Southern Observatory, Karl-Schwarzschild-Strasse 2, D-85748 Garching bei M\"unchen, Germany}

\author[0000-0002-2058-5670]{Karen J. Meech}
\affiliation{Institute for Astronomy, 2680 Woodlawn Drive, Honolulu, HI 96822 USA}

\author[0000-0001-6194-3174]{Beatrice E. A. Mueller}
\affiliation{Planetary Science Institute, 1700 East Fort Lowell, Suite 106, Tucson, AZ 85719-2395}

\author[0000-0002-4734-8878]{Jan T. Kleyna}
\affiliation{Institute for Astronomy, 2680 Woodlawn Drive, Honolulu, HI 96822 USA}

\author[0000-0003-0951-7762]{Harold A. Weaver}
\affiliation{Johns Hopkins University, Bloomberg 145, APL 200-E210, 3400 N. Charles Street, Baltimore MD 21218 USA}

\author[0000-0003-0854-745X]{Marc W. Buie}
\affiliation{Southwest Research Institute, 1050 Walnut St., Suite 300, Boulder, CO 80302 USA}

\author[0000-0001-5159-2104]{Micha\l{} Drahus}
\affiliation{Astronomical Observatory, Jagiellonian University, ul. Orla 171, 30-244, Krak\'{o}w, Poland}

\author[0000-0002-5019-4056]{Piotr Guzik}
\affiliation{Astronomical Observatory, Jagiellonian University, ul. Orla 171, 30-244, Krak\'{o}w, Poland}

\author[0000-0002-1341-0952]{Richard J. Wainscoat}
\affiliation{Institute for Astronomy, 2680 Woodlawn Drive, Honolulu, HI 96822 USA}

\author[0000-0001-6287-7332]{Wac\l{}aw Waniak}
\affiliation{Astronomical Observatory, Jagiellonian University, ul. Orla 171, 30-244, Krak\'{o}w, Poland}

\author[0000-0001-7452-7647]{Barbara Handzlik}
\affiliation{Astronomical Observatory, Jagiellonian University, ul. Orla 171, 30-244, Krak\'{o}w, Poland}

\author[0000-0002-1557-0343]{Sebastian Kurowski}
\affiliation{Astronomical Observatory, Jagiellonian University, ul. Orla 171, 30-244, Krak\'{o}w, Poland}

\author[0000-0002-8808-4282]{Siyi Xu}
\affiliation{Gemini Observatory, 670 N. A'ohoku Place, Hilo HI, 96720 USA}

\author[0000-0003-3145-8682]{Scott S. Sheppard}
\affiliation{Carnegie Institution for Science, 5241 Broad Branch Rd. NW, Washington, DC 20015 USA}

\author[0000-0001-7895-8209]{Marco Micheli}
\affiliation{ESA SSA-NEO Coordination Centre, Largo Galileo Galilei, 1, 00044 Frascati (RM), Italy}
\affiliation{INAF - Osservatorio Astronomico di Roma, Via Frascati, 33, 00040 Monte Porzio Catone (RM), Italy}

\author[0000-0001-8429-2739]{Harald Ebeling}
\affiliation{Institute for Astronomy, 2680 Woodlawn Drive, Honolulu, HI 96822 USA}

\author[0000-0002-2021-1863]{Jacqueline V. Keane}
\affiliation{Institute for Astronomy, 2680 Woodlawn Drive, Honolulu, HI 96822 USA}

\begin{abstract}

We show that \Ou's excited spin could be in a high energy LAM state, which implies that its shape could be far from the highly elongated shape found in previous studies. CLEAN and ANOVA algorithms are used to analyze  \Ou's lightcurve using 818 observations over 29.3~days. Two fundamental periodicities are found at frequencies (2.77$\pm$0.11) and (6.42$\pm$0.18)~cycles/day, corresponding to (8.67$\pm$0.34)~h and (3.74$\pm$0.11)~h, respectively. The phased data show that the lightcurve does not repeat in a simple manner, but approximately shows a double minimum at 2.77~cycles/day and a single minimum at 6.42~cycles/day. This is characteristic of an excited spin state. \Ou\ could be spinning in either the long (LAM) or short (SAM) axis mode. For both, the long axis precesses around the total angular momentum vector with an average period of (8.67$\pm$0.34)~h. For the three LAMs we have found, the possible rotation periods around the long axis are 6.58, 13.15, or 54.48~h, with 54.48~h being the most likely. \Ou\ may also be nutating with respective periods of half of these values. We have also found two possible SAM states where \Ou\ oscillates around the long axis with possible periods at 13.15 and 54.48~h, the latter as the most likely. In this case any nutation will occur with the same periods. Determination of the spin state, the amplitude of the nutation, the direction of the TAMV, and the average total spin period may be possible with a direct model fit to the lightcurve. We find that \Ou\ is ``cigar-shaped'', if close to its lowest rotational energy, and an extremely oblate spheroid if close to its highest energy state for its total angular momentum. 
\end{abstract}



\section{Introduction} \label{sec:intro}

\begin{figure*}[ht!]
\plotone{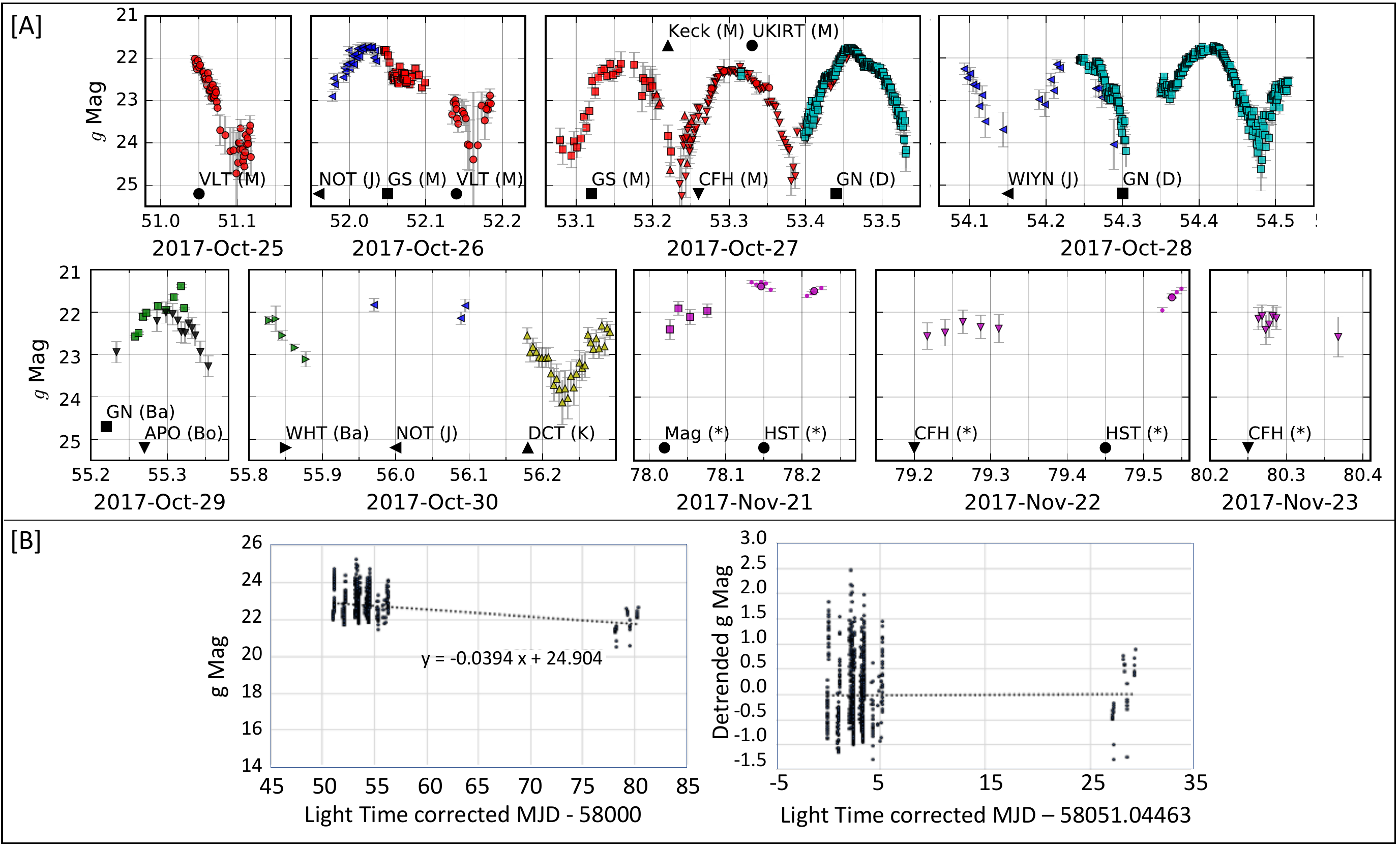}
\caption{[A] Photometric data used for this study, converted to $g$ band, corrected for geometry and light-travel time to 2017~Oct.~25. The epochs are in (JD$-245800.5$). The provenance of the data is as follows: 
*:~this paper;
Ba:~\citet{bannister2017}; 
Bo:~\citet{bolin2017}; 
D:~\citet{drahus2017}; 
J:~\citet{jewitt2017}; 
K:~\citet{knight2017}; 
M:~\citet{meech2017}. The colors and symbols differentiate the data sources.
[B] Left: Data reduced to $g$ magnitudes. Right: $g$ data with linear trend removed and time reduced to zero for the first observation point. These ``detrended'' data are the basis for the frequency analysis.}
\label{fig:data}
\end{figure*}

The lightcurve of the interstellar object 1I/2017 U1 (\Ou) has been the subject of intense series of observations to determine, among other properties, its rotation period \citep{meech2017, bolin2017, bannister2017, drahus2017, feng2017, fraser2017, jewitt2017, knight2017}. Several of these authors have noted that the lightcurve showed the characteristics of an excited or `tumbling' motion \citep{fraser2017, drahus2017} but did not further pursue a detailed analysis; other authors \citep{meech2017, bolin2017, jewitt2017}  analyzed their data sets in terms of a simple rotator. All of these authors offered estimates of the rotation period, which varied between 6.9 and 8.3~h, under the assumption of a double-peaked phase curve, characteristic of an elongated object with little or no albedo contrast on its surface. In this paper we analyze most of the published and shared observations of the lightcurve. The 818 observations, spanning a time interval of 29.3~days, show that there are two dominant and several related compound frequencies in the lightcurve frequency spectrum, which allow several, but not all, important properties of the rotation state to be determined. In particular, we show that \Ou\ may be in a high energy state, which has important implications for its shape.

\section{Construction of the Lightcurve} 

\begin{table*}
\caption{\label{tab:geometry}Observing Geometry}
\begin{center}
\begin{tabular}{cccccccll}
\hline
\hline
\multicolumn{2}{c}{Begin UT Date, MJD$^{\dag}$} &\multicolumn{2}{c}{End UT Date, MJD$^{\dag}$} & $r$$^{\ddag}$ & $\Delta$$^{\ddag}$ & $\alpha$$^{\ddag}$ & Telescope & Reference\\
& & & & [au] & [au] & [deg] & & \\
\hline
Oct 25 01:04 & 51.045 & Oct 25 02:49 & 51.118 & 1.361 & 0.399 & 19.3  & VLT & \citet{meech2017} \\
Oct 25 23:28 & 51.978 & Oct 26 00:50 & 52.035 & 1.384 & 0.430 & 20.7 & NOT & \citet{jewitt2017} \\
Oct 26 01:05 & 52.046 & Oct 26 02:25 & 52.101 & 1.386 & 0.431 & 20.8 & Gemini S& \citet{meech2017} \\
Oct 26 03:12 & 52.134 & Oct 26 04:26 & 52.185 & 1.388 & 0.434 & 20.9 & VLT & \citet{meech2017} \\
Oct 27 01:51 & 53.078 & Oct 27 05:24 & 53.226 & 1.411 & 0.467 & 22.1 & Gemini S & \citet{meech2017} \\
Oct 27 05:39 & 53.236 & Oct 27 10:57 & 53.457 & 1.416 & 0.473 & 22.3 & CFHT & \citet{meech2017} \\
Oct 27 05:48 & 53.242 & Oct 27 06:01 & 53.251 & 1.413 & 0.471 & 22.2 & Keck & \citet{meech2017} \\
Oct 27 07:34 & 53.316 & Oct 27 12:46 & 53.532 & 1.417 & 0.477 & 22.4 & Gemini N &  \citet{drahus2017} \\
Oct 28 02:14 & 54.094 & Oct 28 06:56 & 54.289 & 1.436 & 0.503 & 23.1 & WIYN &  \citet{jewitt2017} \\
Oct 28 05:52 & 54.245 & Oct 28 12:25 & 54.518 & 1.441 & 0.509 & 23.3 & Gemini N & \citet{drahus2017} \\
Oct 29 05:37 & 55.234 & Oct 29 08:30 & 55.354 & 1.463 & 0.541 & 24.0 & APO & \citet{bolin2017} \\
Oct 29 06:12 & 55.259 & Oct 29 07:44 & 55.323 & 1.462 & 0.540 & 24.0 & Gemini N & \citet{bannister2017} \\
Oct 29 19:52 & 55.828 & Oct 29 21:04 & 55.878 & 1.475 & 0.560 & 24.4 & WHT &  \citet{bannister2017} \\
Oct 29 23:18 & 55.971 & Oct 30 03:17 & 56.137 & 1.480 & 0.567 & 24.6 & NOT & \citet{jewitt2017} \\
Oct 30 04:19 & 56.180 & Oct 30 07:01 & 56.293 & 1.485 & 0.573 & 24.7 & DCT & \citet{knight2017} \\
Nov 21 00:46 & 78.033 & Nov 21 01:57 & 78.082 & 1.980 & 1.364 & 27.2 & Magellan & This paper \\
Nov 21 03:20 & 78.140 & Nov 21 05:32 & 78.231 & 1.983 & 1.370 & 27.2 & HST & This paper \\
Nov 22 05:07 & 79.214 & Nov 22 07:48 & 79.326 & 2.007 & 1.409 & 27.1 & CFHT & This paper \\
Nov 22 12:43 & 79.530 & Nov 22 13:19 & 79.555 & 2.012 & 1.421 & 27.0 & HST & This paper \\
Nov 23 06:28 & 80.270 & Nov 23 09:11 & 80.383 & 2.029 & 1.450 & 26.9 & CFHT & This paper \\
\hline
\end{tabular}
\end{center}

\vspace{0.2cm}
{\bf Notes:} $^{\dag}$Epoch of first and last exposures of each run, in UT and MJD = JD-2458000.5; $^{\ddag}$$r, \Delta$: helio- and geocentric distances, $\alpha$: solar phase angle (from Horizon ephemerides JPL\#10).
\end{table*}

\begin{table}
\caption{\label{tab:mag}New Observations}

\begin{tabular}{cccccc}
\hline
\hline
2017 Nov & mJD & Mag$^{\dag}$ & $\sigma$$^{\dag}$ & Filter & Telescope \\
\hline
21 00:46 & 78.033 & 25.42 & 0.24 & w & Magellan \\
21 01:03 & 78.044 & 24.92 & 0.16 & w &  \\
21 01:25 & 78.059 & 25.13 & 0.18 & w &  \\
21 01:57 & 78.082 & 24.99 & 0.16 & w &  \\
21 03:20 & 78.140 & 24.79 & 0.04 & V & HST \\
21 03:29 & 78.146 & 24.84 & 0.03 & V &  \\
21 03:38 & 78.152 & 24.78 & 0.03 & V &  \\
21 03:47 & 78.158 & 24.82 & 0.03 & V &  \\
21 03:56 & 78.165 & 24.96 & 0.04 & V &  \\
21 05:05 & 78.212 & 25.11 & 0.04 & V &  \\
21 05:14 & 78.218 & 25.05 & 0.04 & V &  \\
21 05:32 & 78.231 & 24.93 & 0.04 & V &  \\
22 05:20 & 79.223 & 25.67 & 0.31 & w & CFHT \\
22 05:54 & 79.246 & 25.59 & 0.32 & w &  \\
22 06:27 & 79.269 & 25.34 & 0.28 & w &  \\
22 07:01 & 79.293 & 25.46 & 0.28 & w &  \\
22 07:35 & 79.316 & 25.50 & 0.33 & w &  \\
22 12:43 & 79.530 & 25.56 & 0.06 & V & HST \\
22 13:01 & 79.543 & 25.21 & 0.04 & V &  \\
22 13:10 & 79.549 & 25.13 & 0.04 & V &  \\
22 13:19 & 79.555 & 25.05 & 0.04 & V &  \\
23 06:28 & 80.270 & 25.34 & 0.26 & w & CFHT \\
23 06:35 & 80.274 & 25.28 & 0.25 & w &  \\
23 06:41 & 80.279 & 25.60 & 0.35 & w &  \\
23 06:48 & 80.284 & 25.48 & 0.31 & w &  \\
23 06:55 & 80.288 & 25.28 & 0.25 & w &  \\
23 07:01 & 80.293 & 25.33 & 0.26 & w &  \\
23 08:58 & 80.374 & 25.78 & 0.47 & w &  \\
\hline
\end{tabular}

\vspace{0.2cm}
{\bf Notes:} Mid-exposure epochs (in UT, and MJD=JD-2458000.5), uncorrected for light travel time; $^{\dag}$Magnitude uncorrected for geometry and $1\sigma$ error. 
\end{table}

\label{sec:observations}

\subsection{Published data}
The observations published in \citet{meech2017, bolin2017, bannister2017, drahus2017, fraser2017, jewitt2017} and \citet{knight2017} have been collected and converted to the $g$-band using the transformations listed in \citet{jordi2006} with the colors published in these respective papers or in \citet{meech2017} where needed. In the case of the CFHT wide $gri$ filter, the color conversion from \citet{tonry2012} was used.

\subsection{Additional data}
We obtained additional images on the nights of 2017 November 22 and 23 using the CFHT MegaCam imager, an array of forty 2048$\times$4612 pixel CCDs with a plate scale of 0$\farcs$187 per pixel and a 1.1 square degree FOV.  The data were obtained through the wide $w$ ($gri$-band) filter, using service observing with the telescope guided at non-sidereal rates during exposures of 360 seconds.  The images were processed through the Elixir pipeline \citep{magnier2004} to remove the instrumental signature.  

The Magellan-Baade 6.5 meter telescope in Chile at Las Campanas Observatory observed the object on 2017 November 21, 22 and 23 with the wide-field IMACS camera, which has eight $2048\times4096$ pixel CCDs with 0$\farcs$20 per pixel. The nights were photometric with seeing between 0$\farcs$6 and 0$\farcs$8. The object was imaged through the broad WB4800-7800 filter, which transmits most of the light between 0.480-0.780~$\mu$m to the detector.  Biases and dithered twilight flats were used to calibrate the CCDs.  The telescope was tracked at non-sidereal rates during exposures of 450 to 600~s.

We processed the CFHT and Magellan data using the same technique and tools as described in \citet{meech2017}: we use the Terapix/Astromatic tools \citep{bertin1996} to fit world coordinates (RA and Dec) based on reference stars from the SDSS and 2MASS catalogs. We used expanded SExtractor \citep{bertin1996} automatic apertures to measure the magnitudes of trailed stars and computed a photometric zero point for each image based on stars from the PS1 database \citep{magnier2016} 3-pi survey \citep{chambers2016} or the Sloan Digital Sky survey \citep{fukugita1996}. The $w$-band filter and the WB4800-7800 filters were converted to $g$-band using the colors reported in \citet{meech2017}.

Series of images were acquired with the Hubble Space Telescope using the UVIS channel of the Wide-Field Camera 3 (WFC3) and the F350LP filter. These images were grouped in two orbits on 2017 November~21 and one on November~22, each one including five individual images. \Ou\ was contaminated by cosmic rays in three images out of the fifteen, and photometry is not reported for those cases. 
The raw counts were measured in a circular aperture of 5-pixel (0$\farcs$2) radius, and the background was estimated using an annulus between 10-20 pixels. The raw counts were converted to the standard V-mag (Johnson system) by comparing the observed count rates in a 0$\farcs$2 radius aperture to the count rate predicted to be in that aperture by the WFC-UVIS exposure time calculator assuming a target with a solar spectrum reddened by 23\% per 100 nm \citep{meech2017}. These $V$ magnitudes were then converted to $g$ magnitudes. 

The geometry of all the observations is detailed in Table~\ref{tab:geometry}, and the epoch and magnitudes of the new ones are listed in Table~\ref{tab:mag}.

\subsection{Data Reduction}  \label{sec:reduction}
All the published and new data, converted to $g$ magnitudes, have been scaled to the geometry of 2017 October~25 at 2~UT 
($r=1.3616$~au, $\Delta=0.3983$~au and $\alpha=19.310^{\circ}$, helio- and geocentric distances, and solar phase angle, from orbit JPL\#10). The solar phase effect was corrected using a linear function ($-0.04$~mag/deg, the canonical value for cometary and D-class objects). 

The final data set is shown in Fig.~\ref{fig:data}. Over the full time-span, the data show a weak trend to brighter magnitudes that is likely due to the changing viewing geometry relative to the rotation pole, and to an imperfect correction of the phase effect. To minimize the effect of the mean value of the data and its overall slope on the frequency spectrum, we linearly detrend the data with the regression 
\begin{equation}
g = -0.0394  \;t + 22.892
\end{equation}
where $t$ is the epoch of observations (corrected for light-travel time) minus $2458051.54463$, the epoch of the first point. 
As the phase angle varies monotonically with time over all but the last observation, changing the phase correction will introduce a time-dependent shift in magnitude, which is (partly) corrected by the de-trending. The trend could not be corrected by changing the phase parameter, so other effects must dominate it, and it cannot be used to constrain the phase parameter.

Using these ``detrended'' data in the frequency analysis removes strong responses (and their spectrum of aliases) at zero frequency and at low frequencies associated with the overall time-span of the data. This improves the identification of responses associated with rotation in the resulting frequency spectrum. The detrended data are shown in the right panel of Fig.~\ref{fig:data}B.  Some runs show a systematic deviation with respect to neighboring data, suggesting an issue with the photometric calibration, or with the color conversion (possibly caused by color variations across the object), or with the measurement method (in particular with the aperture correction used for faint objects), or a combination of these and other effects. The change of viewing geometry over the span of the observations (about $14^{\circ}$) could introduce a change in the observed lightcurve timing. However, this effect is small: in the worst case scenario (a rotation axis perpendicular to the great circle tangential to the track of the object), this effect would be of less than 15~min for a 7~h rotation period.

\section{Frequency Analysis}
\label{sec:analysis}

The detrended data are analyzed for temporal frequencies using the CLEAN \citep{belton1988} and ANOVA \citep{sc1996} algorithms. CLEAN was designed to remove alias patterns associated with the prime frequency responses in the spectrum. From a ``dirty'' spectrum, essentially the discrete Fourier transform of the data, a representation of the alias pattern \citep[Spectral Window,][]{deeming1975}, derived from the sampling pattern, is iteratively applied to the most prominent peaks in the spectrum and subtracted until the aliases are effectively removed. It has been used with considerable success on lightcurves of comet 1P/Halley, Toutatis, and several other objects \citep{mueller2002}. ANOVA, part of the Peranso software package\footnote{www.CBABegium.com} and efficient at damping aliases in the frequency spectrum, provides a powerful analysis-of-variance algorithm that has been successfully used on comets 103P/Hartley 2 and 9P/Tempel 1 spacecraft data \citep{belton2011, belton2013}. We used ANOVA with a 3-harmonic basis, which gives the strongest frequency responses aligned with CLEAN even though it shows stronger aliasing than the more often used 2-harmonic basis. With the latter, the alias pattern is  less confused, but then the frequency of the peak responses lead to conflicts with CLEAN, even though the phase plots associated with the 2-harmonic ANOVA peaks show improved order. These problems show the value of using multiple algorithms to come to a conclusion in this kind of analysis. While some of the individual runs display a systematic magnitude offset with respect to neighboring data, their frequency information is unaffected. This was checked by repeating the analysis omitting each affected run, and verifying that the results are not changed. 

\begin{figure}[ht!]
\plotone{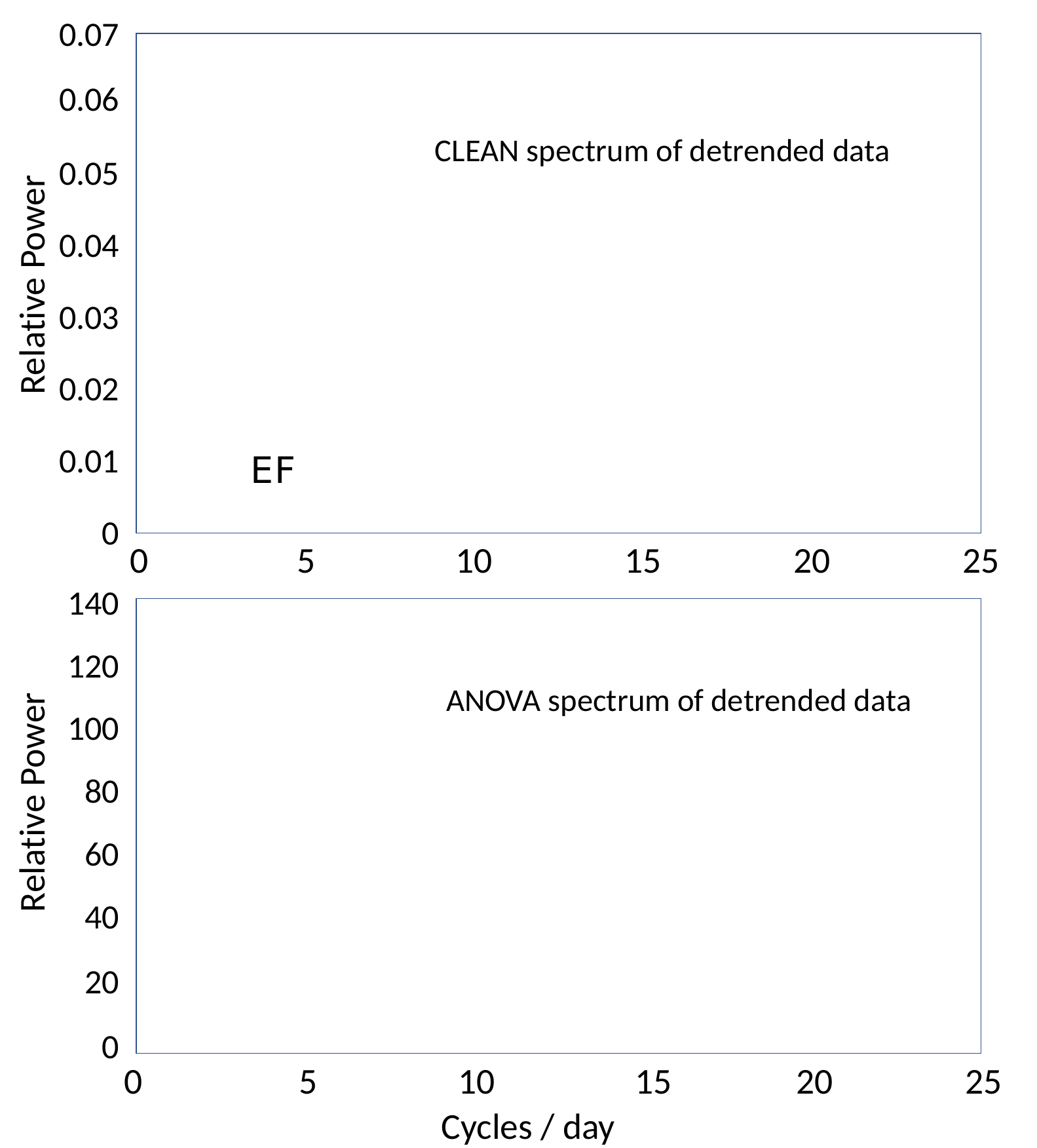}
\caption{Frequency spectrum of the detrended data using the CLEAN and ANOVA algorithms. The peaks at A and C are of primary interest because they are clearly present in the spectra of both algorithms. The peaks at B, D, E and F are discussed in the text.}
\label{fig:frequency}
\end{figure}

The results are shown in Fig.~\ref{fig:frequency}, which displays spectra of the detrended data out to 25~cycles/day. Most of the power is at frequencies of less than 10~cycles/day, and the very low noise level can be judged from the spectra within the interval from 20 to 25~cycles/day. Both the ANOVA and CLEAN spectra contain two unrelated features (A, C), as expected for a body in an excited rotation state, at essentially the same frequencies. The dominant frequencies of these features are  (2.77$\pm$0.11)~cycles/day (A) and ($6.42\pm0.18$)~cycles/day (C), corresponding to periodicities of ($8.67\pm0.34$)~h and ($3.74\pm0.11$)~h, respectively.  The frequency of B (5.65~cycles/day, 4.25~h period) is twice that of A, suggesting a clear relationship. We also note that C is at twice the rotational frequency of 3.18~cycles/day found by \citet{drahus2017}. However, while this frequency is consistent with a double-minimum phase curve and the single minimum in the phase curve at C (see below), no spectral peak is present near 3.18~cycles/day in our CLEAN spectrum, suggesting that C is a compound frequency response. The peak at D (0.31~cycles/day, 3.226~day period) is probably unrelated to rotation and may be the result of the extent of the large data sample between 0 and 5~days and the large time gaps in the sampling of the data (Fig.~\ref{fig:data}B). 

\section{Interpretation}
\label{sec:interpretation}

Our basic assumptions are that \Ou\ is a single object, and that it rotates as a rigid body free of torques. The assumption of rigidity is not completely assured if the object is a rubble pile or extremely weak. Nevertheless, experience has shown that this is a reasonable assumption for cometary nuclei and small asteroids and may apply to \Ou. Our assumption that the object is free of torques is based on observations by \citet{meech2017, knight2017, jewitt2017, ye2017, drahus2017} who find no evidence for activity in deep images of the vicinity surrounding \Ou, and on deep spectra by \citet{fitzsimmons2017} showing no cometary emission lines. Other possible torques (e.g. solar radiation pressure) are expected to be extremely weak and unlikely to affect the motion during the object’s short fly-through of the solar system.

\begin{figure}[ht!]
\plotone{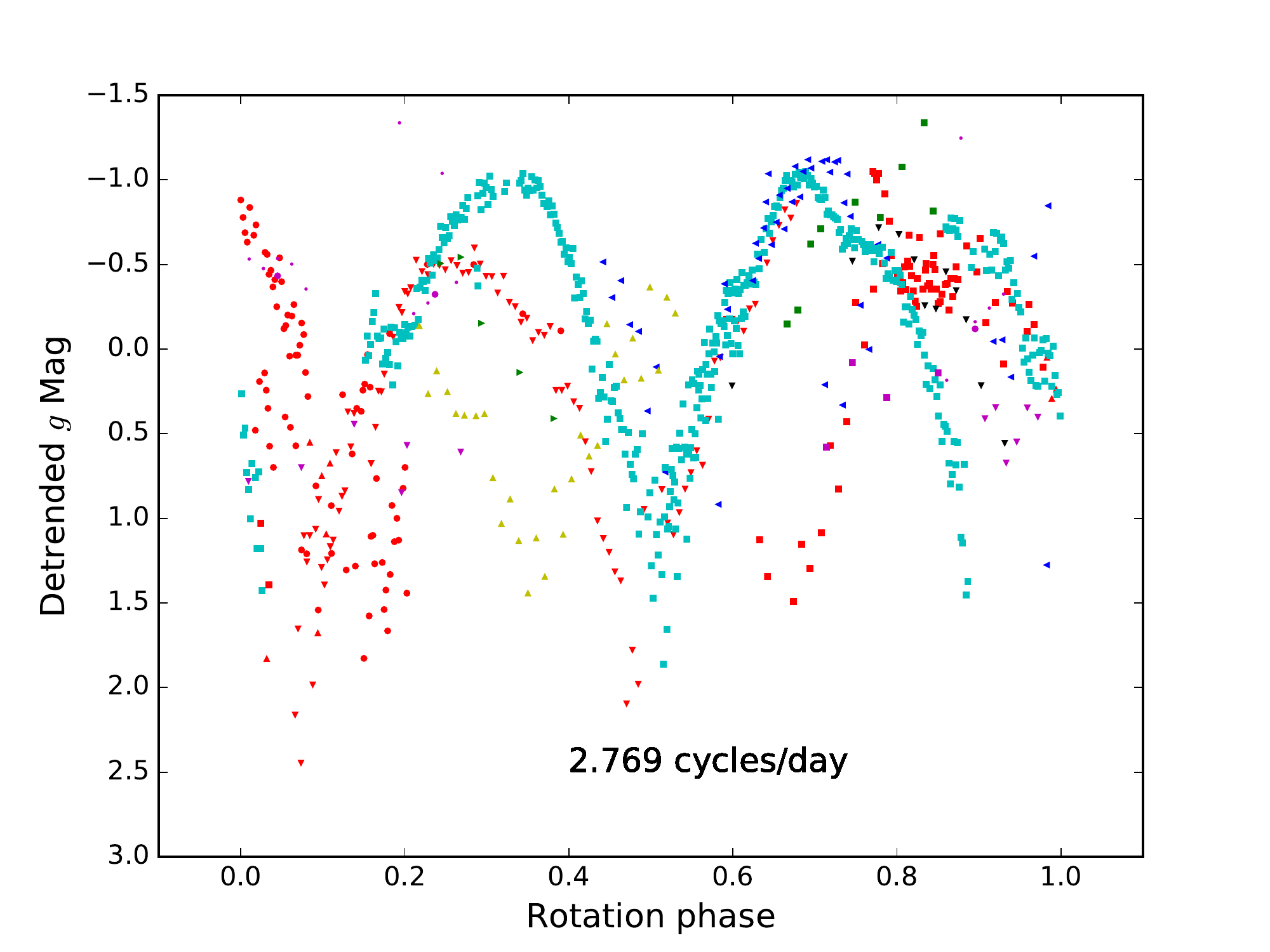}
\plotone{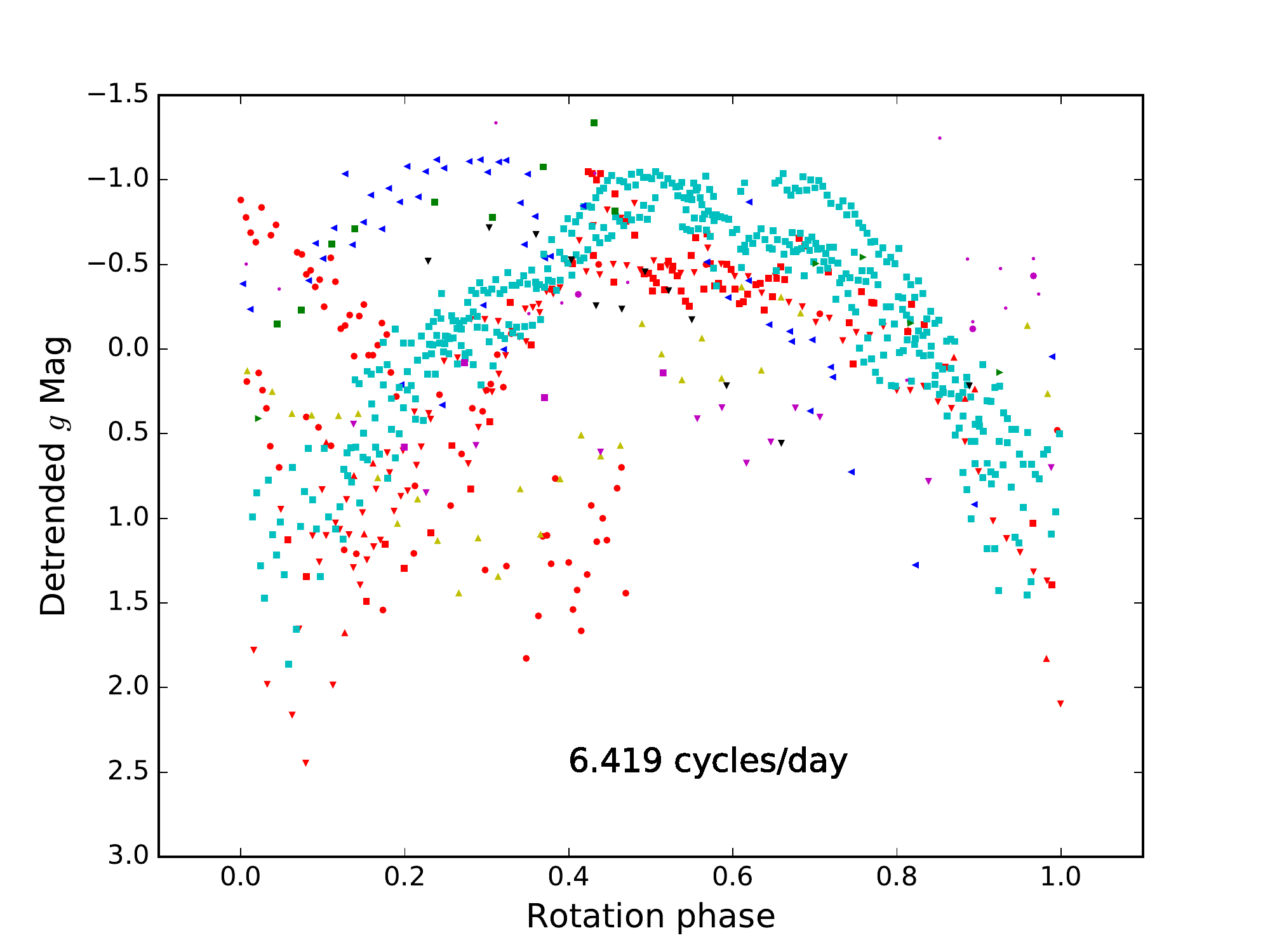}
\caption{Phase curves for frequencies A (upper panel) and C (lower panel) in the spectra of the detrended data. The symbols are the same as in Fig.~1A.
}
\label{fig:phase}
\end{figure}

Phase curves for A and C are shown in Fig.~\ref{fig:phase}. As expected for rotation in an excited state the curves do not repeat well. This is because the body does not generally return to the same geometric orientation with respect to the line-of-sight (LOS) after a complete precession of its long axis around the total angular momentum vector (TAMV). We include the plots because they can be diagnostic of the type of spin state and useful in interpreting the primary frequencies in the spectra. The phase plot of A shows two minima per cycle, while that of C has a single minimum.

\citet[hereafter SM15]{samarasinha2015} have shown that the number of strong responses expected in the frequency spectrum of a non-principal rotator depends on the degree of excitation of the rotation state. In particular, the responses will be different for spin in the Short Axis Mode (SAM) and the Long Axis Mode (see SM15 for definitions).
The number and relative strength of the responses also depend on the shape of the object (assumed to have no large albedo variations) and on the angle between the LOS and the space direction of the TAMV. The number of strong responses is usually less than 5. We take the listing from SM15 (Table 2) for the origin of the six most probable frequency responses as the basis for our analysis. To define the motion, we use the L-convention as defined in SM15. This means that the Euler angles $\phi$, $\psi$, $\theta$ are referring to the long axis of the body as it precesses, rotates (or oscillates), and nutates (up-and down nodding of the long axis) around the TAMV. The periods associated with these component motions are $P_{\phi}$, $P_{\psi}$, $P_{\theta}$. 

Given the extreme elongation of \Ou\ \citep{meech2017} a double minimum in the lightcurve is expected per precessional cycle of the long axis around the TAMV. This identifies either A or C/2 as the frequencies associated with $P_{\phi}$. The probable compound rotational signatures listed by SM15 (Table 2) then allow the determination of $P_{\psi}$. We find $P_{\psi}< 0$ for $P_{\phi} = 2/C$ in all cases, which is not allowed for either LAMs or SAMs (\citet{samarasinha1991}, Appendix; hereafter SA91). In addition, and as noted above, no spectral response is found at frequency C/2. Therefore, $P_{\phi} = $1/A$ = (8.67\pm 0.34$)~h. Since B is 2A, it follows that B is a response at 2/$P_{\phi}$. For the probable compound frequencies listed in SM15, $1/P_{\phi}+ 1/P_{\psi}$, $2/P_{\phi} + 2/P_{\psi}$, and $1/P_{\phi} + 2/P_{\psi}$, the allowed values for $P_{\psi}$ are 6.58, 13.15, and 54.48~h. To choose between these possibilities we consider the two peaks in the CLEAN spectrum at 3.44 (E) and 4.18~cycles/day (F). The first of these can be satisfied by compound responses involving any of the allowed values of $P_{\psi}$ and is thus not diagnostic. However, the response at 4.18~cycles/day only appears to be satisfied by compound frequencies at 1/$P_{\phi}$+3/$P_{\psi}$ and 2/$P_{\phi}$--3/$P_{\psi}$ for $P_{\psi}$ = 54.48~h. This period is therefore our most likely estimate of the roll or oscillatory period around the long axis. While they are not in the SM15 list, these frequencies are of course possible, as might be other compound frequencies not yet identified. Without a direct fit to the lightcurve, the choice of $P_{\psi}$ remains nevertheless uncertain. Any of the allowed values of $P_{\psi}$ noted above are possible LAM states. However, since $P_{\psi} / P_{\phi}< 1$ for $P_{\phi} = 6.58$~h, this cannot be a SAM state (SA91). For the two SAM states that remain, $P_{\theta} = P_{\psi}$, while for the three possible LAM states, $P_{\theta} = P_{\psi} / 2$ (SA91).

\section{Discussion}
\label{sec:discuss}

\Ou\ is in an excited rotational state with its long axis  irregularly precessing around the TAMV with an average period of (8.67$\pm$0.34)~h. It is also nutating, unless it is a symmetric rotator ($b = c$) in a LAM state, in which case it is required to precess at a constant rate around the TAMV inclined at a constant angle $\theta$ (SA91).
It may seem odd that the very strong spectral response at C is a compound of $P_{\phi}$ and $P_{\psi}$ and not simply related to $P_{\phi}$, but this is not unusual (see, for instance, Case 1 of Figs. 5 and 6 in SM15) and may possibly also be the reflection of a shape that is far from symmetric. \Ou\ also rotates around its long axis. Whether this motion is a complete rotation (LAM), or an oscillation (SAM), is not determined. However, in the case of a LAM, the likely possible periods associated with this motion are 6.58, 13.15, and 54.48~h. In the case of a SAM, the possible periods are 13.15, and 54.48~h. Our best, but nevertheless uncertain, estimate of the most likely roll or oscillatory period is 54.48~h. The amplitudes of any oscillation or nutation are not determined. 
These results are only based on the temporal frequency spectrum, using the probable compound frequencies resulting from the general analysis of excited rotation (from SM15 Table 2), complemented by two additional ones associated to peak F at 4.18~cycles/day.

The values of $P_{\phi}$ and $P_{\psi}$ found here also place constraints on the shape of \Ou, approximated here by an ellipsoid with a$>$b$>$c. \citet{meech2017} have already shown that the object is highly elongated. However, the periods $P_{\phi}$ and $P_{\psi}$,  through equations A53 and A80 in SA91, place additional limits on $b/a$ (if a LAM) or $c/b$ (if a SAM). We find that these new limits show, in the case of a SAM, that $b$ can be at most 1.7 times longer than $c$, and, in the most likely case ($P_{\psi}$~=~54.48~h), at most 1.03 times longer than $c$, i.e., the object is crudely ``cigar'' shaped. This is the case considered in \citet{meech2017}, who effectively assumed a minimal amount of rotational energy in the spin state. In the case of the LAM, we find that $b$ can be much larger than $c$. With $P_{\psi} = 6.58$~h, $0.1 < b/a < 0.7$; $P_{\psi} = 13.15$~h, $0.1 < b/a < 0.85$; and for $P_{\psi} = 54.48$~h, the most likely case, $0.1 < b/a < 0.98$. This means that, if \Ou\ is rotating in a LAM state, its shape could be anything from ``cigar-like'' to approximately ``pancake-like'' (a highly oblate ellipsoid rotating around one of its diameters). A LAM state, which is so far not precluded by the observations, includes the case in which the rotational energy is close to maximal (i.e., the instantaneous spin vector is more closely aligned with the long axis) and the shape would need to be an extremely oblate spheroid. Note that if inertias around $b$ and $c$ are equal (a symmetric rotator), then the object, if in an excited state, must spin in the LAM state with no nutation.

Further advances of our knowledge of the rotational state will require using the allowed periods found here to iteratively model the full lightcurve, while varying the shape of the object and the orientation of the TAMV. We are in the process of attempting to extend the time-span of the data and expect to address the significant problem of a direct fit to the full lightcurve in the near future.

It is interesting to contemplate the implications of an excited rotation state for \Ou\ in terms of the timescales for ejection from it's host planetary system.  Several analyses suggest that \Ou\ may have been recently ejected from its host system \citep{feng2017, gaidos2017}. However, as no host star system has been identified yet \citep{feng2017, ye2017, dy2017, zhang2018, zuluaga2017, zwart2017}, it is possible that \Ou\ has been traveling for a long time. Under this scenario it might be expected that the spin might have relaxed to principal axis rotation. The damping timescale from an excited rotation to a state of principal axis rotation around the axis of maximum moment of inertia is given by
\begin{equation}
\tau \sim \frac{\mu Q}{\rho K_3^2 r_n^2 \omega^3}
\end{equation}
where $\mu$ is the material rigidity, $Q$ is the ratio of the oscillation energy to the energy lost each cycle, $\rho$ is the bulk density, $K_3^2$ is a numerical factor relating to the body elongation, $r_n$ the body's average radius, and $\omega$ is the rotation angular frequency \citep{burns1973}. Using values of $K_3^2$~=~0.1 (for highly elongated bodies) and $\mu$Q~=~5$\times$10$^{11}$ N~m$^{-2}$ representative of small solar system bodies \citep{harris1994}, a radius of $r_n$~=~102~m \citep{meech2017}, and densities ranging from cometary to planetary (0.5 $< \rho <$ 6 kg/m$^3$) gives $\tau$~$>$~10$^{11}$ yr.  Thus, the excited spin state of \Ou\ may reflect the processes that ejected it from its home planetary system.

{\it Acknowledgements}
KJM and JTK acknowledge support through awards from the National Science Foundation AST1413736 and AST1617015. RJW acknowledges support by the National Aeronautics and Space Administration under grant NNX14AM74G issued through the SSO Near Earth Object Observations Program. The contributions of MJSB to this research were made pro bono.

Based also in part on observations obtained with MegaPrime/MegaCam (a joint project of CFHT and CEA/DAPNIA, at the Canada-France-Hawaii Telescope which is operated by the National Research Council of Canada, the Institute National des Science de l'Univers of the Centre National de la Recherche Scientifique of France, and the University of Hawai'i), with the 6.5 m Magellan Telescopes (located at Las Campanas Observatory, Chile), and with the NASA/ESA {\sl Hubble Space Telescope}, obtained at the Space Telescope Science Institute (which is operated by the Association of Universities for Research in Astronomy, Inc., under NASA contract NAS 5-26555; these observations are associated with GO program 15405). We are extremely grateful to the second anonymous referee, whose comments and suggestions were extremely helpful and valuable.

\end{document}